\documentclass[iop]{emulateapj}
\usepackage{graphicx}
\usepackage{natbib}
\usepackage{ulem}
\usepackage[usenames]{color}
\usepackage{enumitem}
\usepackage{epstopdf}
\slugcomment{~~\today}

\newcommand       \msun        	{$M_{\odot}$}
 
\newcommand       \lsun      	{$L_{\odot}$} 
\newcommand       \kms          {km~s$^{-1}$}

\newcommand       \mic        	 {$\mu$m}

\newcommand \hersh		{{\it Herschel}}
\newcommand \sild		{MgSiO$_3$}
\newcommand \silm		{Mg$_2$SiO$_4$}

\begin{document}

\title{THE EVOLUTION OF DUST MASS IN THE EJECTA OF SN1987A}

%Greenbelt, MD 20771, \\ e-mail: eli.dwek@nasa.gov}
\author{Eli Dwek\altaffilmark{1}, Richard G. Arendt\altaffilmark{1,2}}

\altaffiltext{1}{Observational Cosmology Lab, Code 665, NASA Goddard Space Flight Center, Greenbelt, MD 20771, USA; e-mail: eli.dwek@nasa.gov}
\altaffiltext{2}{CRESST, University of Maryland -- Baltimore County, Baltimore, MD 21250, USA}

\begin{abstract}
We present a new analysis of the infrared (IR) emission from the ejecta of SN1987A covering days 615, 775, 1144, 8515, and 9090 after the explosion. We show that the observations are consistent with the rapid formation of about 0.4~\msun\ of dust, consisting of mostly silicates (\sild), near day 615, and evolving to about 0.45~\msun\ of composite dust grains consisting of $\sim 0.4$~\msun\ of silicates and $\sim 0.05$~\msun\ of amorphous carbon after day $\sim$~8500. The proposed scenario challenges previous claims that dust in SN ejecta is predominantly carbon, and that it grew from an initial mass of $\sim 10^{-3}$~\msun, to over 0.5~\msun\ by cold accretion. It alleviates several problems with previous interpretations of the data: (1) it reconciles the abundances of silicon, magnesium, and carbon with the upper limits imposed by nucleosynthesis calculations; (2) it eliminates the requirement that most of the dust observed around day 9000 grew by cold accretion onto the $\sim 10^{-3}$~\msun\ of dust previously inferred for days 615 and 775 after the explosion; and (3) establishes the dominance of silicate over carbon dust in the SN ejecta. At early epochs, the IR luminosity of the dust is  powered by the radioactive decay of $^{56}$Co, and at late times  by at least $(1.3-1.6)\times10^{-4}$~\msun\ of $^{44}$Ti.  Even if only a fraction $\gtrsim 10$\% of the silicate dust survives the injection into the ISM, the observations firmly establish the role of core collapse SNe as the major source of thermally condensed silicate dust in the universe.
\end{abstract}

%================================================================
\section{INTRODUCTION}
%================================================================

First evidence for dust emission from the ejecta of SN1987A was provided by high-resolution mid-IR images obtained with the Thermal Region Camera and Spectrograph (T-ReCS) on the Gemini South 8m telescope \citep{bouchet04}. Spatially resolved images of SN1987A at 450 and 870~\mic\ obtained with the Atacama Large Milimmeter/Submillimeter Array (ALMA) between days $\sim 9200$ and $\sim 9300$ after the explosion provided further proof of thermal IR emission from the supernova (SN) ejecta \citep{indebetouw14,zanardo14b}. Combined with almost contemporaneous data obtained on day 9090 after the explosion with the \hersh\ satellite, the infrared (IR) observations have been interpreted as evidence for the presence of a large amount of cold dust \citep{matsuura11,matsuura15}. If the emission arises from only silicates in the form of \silm\, then the observations require 2.4~\msun\ of dust, that is, the precipitation of, respectively,  0.48 and 0.8~\msun, of Si and Mg from the gas phase of the ejecta.  This mass greatly exceed the SN yield of either Si or Mg ($\sim 0.1$~\msun, see Table~\ref{yields} and references therein). The mass of Si required to fit the IR spectrum can be reduced by the inclusion of carbon dust. A combination of silicate and amorphous carbon reduces the mass of silicates to 0.5~\msun. However, it requires an excessive amount of carbon, about (0.3-0.5)~\msun\ or over 2 to 3 times the yield of C, to be locked up in dust. 

The mass of dust inferred from the late time ($t \gtrsim 8500$~d) IR emission also greatly exceeds the $\sim 10^{-3}$~\msun\ of dust derived from early  observations of SN1987A on days 615, 775, and 1144 \citep{moseley89b, wooden93, dwek92c}.   
This difference has been interpreted as evidence for the continuous growth by accretion of refractory elements onto the seed nuclei that were initially formed in the ejecta \cite[][and references therein]{wesson15}. The ejecta temperature after day $\sim 1200$ has dropped below $\sim$~500~K \citep{sarangi13}. An accretion scenario for the evolution of the dust will therefore require most of the growth to take place by cold accretion, so that the final structure of the dust grains will primarily consist of a mantle, loosely bound with surface energies $\lesssim 1$~eV, surrounding a small thermally-condensed core. Such dust grains will not survive the reverse shock expected to propagate into the ejecta \citep{nozawa07}, contrary to the observed $\sim 0.04$~\msun\ of dust that was subjected to and survived the reverse shock in Cas~A, or in the $\sim$10,000 old Sgr East remnant \citep{arendt14, barlow10, rho08,lau15}. 

Also, the absence of the 9.7 and 18~\mic\ emission features in the IR spectra on days 615 and 775 has been interpreted as evidence that the newly-formed dust consisted of mostly carbon dust \citep{wooden93,ercolano07a,ercolano07b,wesson15}. Using the observed spectra of days 615 and 775 and a fully 3D radiative transfer model, \cite{wesson15} concluded that about $(4-8)\times10^{-4}$~\msun\ of carbon condensed in the ejecta of SN1987A by those epochs. In particular, they derived an upper limit of $\sim 10^{-4}$~\msun\ on the mass of silicates dust that could have formed in the ejecta. This low yield is puzzling, considering the preponderance of Si and Mg in the ejecta, and that SNe should be the major source of thermally-condensed dust in the local and early universe \citep{dwek98,todini01,nozawa03,schneider04,dwek07b,zhukovska08a,cherchneff10,calura10,gall11a,gall11c, dwek11a,calura14,dwek14}.  
 
In this paper we present an alternative model for the evolution of dust in the ejecta of SN1987A, and show that the currently observed dust mass in the ejecta on days $\sim 8500-9100$ could already have been present at the earlier epochs. 
The conclusions of \cite{ercolano07a} and \cite{wesson15} represent just one possible, albeit troublesome, scenario for the evolution of the dust mass in SN1987A. 
Large amounts of dust could have been hidden in optically thick regions of the ejecta, a possibility first raised by \cite{lucy89} and \cite{danziger09}, and  now supported by the ALMA observations. 
We emphasize that our scenario for the evolution of dust mass in SN1987A is not unique, but a viable alternative which addresses the difficulties in previous models: (1) the large mass of either Si, Mg, or C that needs to be locked up in dust to account for the ALMA observations; (2) the growth of dust mass by cold accretion, creating loosely bound mantles that will not survive the passage of the reverse shock; and (3) the absence of any significant amount of silicate dust in the ejecta.

Our alternative model has important implications for the composition of dust in the early universe ($z \gtrsim 7$), before carbon-rich AGB stars left the main sequence. Even if only a fraction $\gtrsim 10$\% of the silicate dust survives the injection into the ISM, the observations of SN1987A firmly establish the role of core collapse SNe as the major source of thermally condensed silicate dust in the early universe. Interstellar dust is then silicate-rich, with important observational consequences for galactic extinction corrections and photometric redshift determinations \citep{dwek14}.

The paper is organized as follows. We first present the basic equations for the escape of IR photons from a homogeneous dusty sphere, the dust model used in our fitting procedure (Section~2). In Section~3 we present the fits to the observations, and the evolution of dust mass.
We first start with the analysis of the late-time emission at epochs 8515 and 9090, when the ejecta was optically thin, and construct a model in which the IR emission arises from elongated composite grains consisting of a silicate \sild\ matrix with amorphous carbon (AC) inclusions. Such configuration can arise as a results of the coagulation of silicates and AC grains at late times in the ejecta. The mass of Mg, Si, and C needed to be locked up in the dust is entirely consistent with nucleosynthesis constraints. We then show that the same mass of refractory elements could have already condensed out of the gas at the early epochs (days 615, 775, and 1144). Here the calculations assumed that the ejecta is homogeneous. A clumpy ejecta model is described in Section~4. In Section~5 we calculate the evolution of the IR luminosity, and compare it to the energy input by radioactivity. A summary and the astrophysical implications of our results are presented in Section~6.

%=============================================================================
\section{THE ESCAPE OF INFRARED PHOTONS FROM A \\ HOMOGENEOUS DUSTY SPHERE}
%============================================================================= 

The general equation for the flux emerging from a homogeneous dusty sphere of radius $R$ at distance $D$ is given by:
\begin{equation} %--- eq 1
\label{flux}
S_{\nu}(\lambda) = {4 M_d\ \kappa(\lambda)\over 4 \pi D^2} \ \pi B_{\nu}(\lambda,\ T_d)\ P_{esc}[\tau(\lambda)] \qquad,
\end{equation}
 where $M_d$ is the mass of dust in the sphere, $\kappa(\lambda)$ is the mass absorption coefficient of the dust, and $B_{\nu}(\lambda,\ T_d)$ is the Planck function at wavelength $\lambda$ and dust temperature $T_d$. The function $P_{esc}$ is the probability for an IR photon to escape from the sphere and given by \cite{cox69,osterbrock06}:
\begin{equation}  %--- eq 2
\label{pesc}
P_{esc}(\tau) = {3\over 4 \tau}\ \left[1-{1\over2\tau^2}+({1\over\tau}+{1\over2\tau^2})e^{-2\tau}\right]
\end{equation}
where $\tau$ is the radial optical depth of the sphere given by:
\begin{equation}  %--- eq 3
\label{tau}
\tau(\lambda) = {3\over4}\ \left({M_d\over \pi R^2}\right)\ \kappa(\lambda) \ .
\end{equation}
Limiting values for $P_{esc}(\tau)$ are:
\begin{eqnarray}   %--- eq 4
\label{pesc2}
P_{esc}(\tau)   & \approx & 1 - {3\tau\over 4} + {2 \tau^2\over 5} -{\tau^3\over 6}\,  \qquad {\rm when} \ \tau << 1 \\ \nonumber 
 & \longrightarrow &\ {3\over 4 \tau}  \qquad \qquad \qquad \qquad \ \ \  {\rm when} \  \tau \longrightarrow \infty  \ .
\end{eqnarray}

The escape formalism explicitly takes  the distribution of optical depth in the sphere into account. As $\tau$ becomes very large, $P_{esc}(\lambda) \sim 1/\kappa(\lambda)$ [eq. (\ref{tau}) and (\ref{pesc2})], with the consequence that the spectral signatures of the dust are cancelled in eq. (\ref{flux}). Furthermore, all information on the dust mass in the sphere becomes indeterminate.  

Equation~(\ref{flux}) and (\ref{tau}) can be readily generalized to a dust population characterized by different compositions, masses, and temperatures. 
The flux is then given by the sum over dust components:
\begin{equation} %--- eq 1
\label{flux2}
S_{\nu}(\lambda) = \sum_j\ {4 M_{d,j}\ \kappa_j(\lambda)\over 4 \pi D^2} \ \pi B_{\nu}(\lambda,\ T_{d,j})\ P_{esc}[\widetilde{\tau}(\lambda)] 
\end{equation}
where $\widetilde{\tau}$ is the optical depth in the presence of multiple dust components:
\begin{equation}
\label{tau2}
 \qquad \widetilde{\tau}(\lambda)  =  {3\over 4 \pi R^2}\ \sum_j M_{d,j}\, \kappa_j(\lambda) 
\end{equation}
where $M_{d,j}$, $\kappa_j(\lambda)$, and $T_{d,j}$ are the masses, mass absorption coefficients, and temperatures corresponding to the different grain compositions $j$. 

In the high optical depth limit, $S_{\nu}(\lambda)$ will still be a blackbody, as long as 
all dust components are at the same temperature. If the temperatures differ, then 
the emission features, generated by $\kappa(\lambda)$, and the absorption features, generated by  $P_{esc}(\lambda)$, will not
cancel perfectly, and the total spectrum will still retain an imprint of the individual dust spectra. 
The former case is illustrated by our fit at Day 775, and the latter case by fits at 
Day 615 and 1144, as shown later in Figures ~\ref{spec2} and \ref{spec3}, and Table~\ref{results}.

%------ figure 1
 \begin{figure*}[tbp]
  \centering
    \includegraphics[width=3.2in]{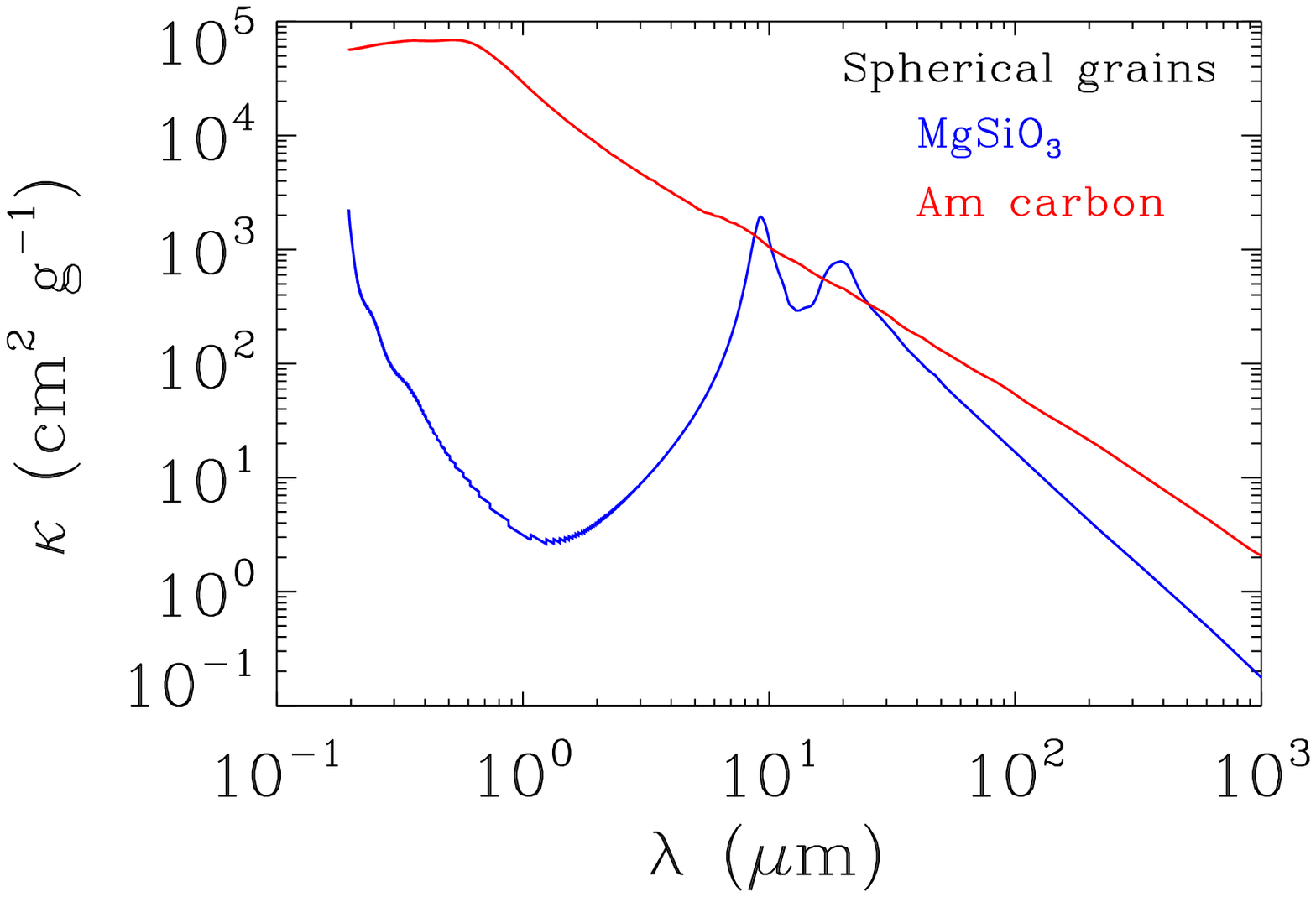} 
  \includegraphics[width=3.2in]{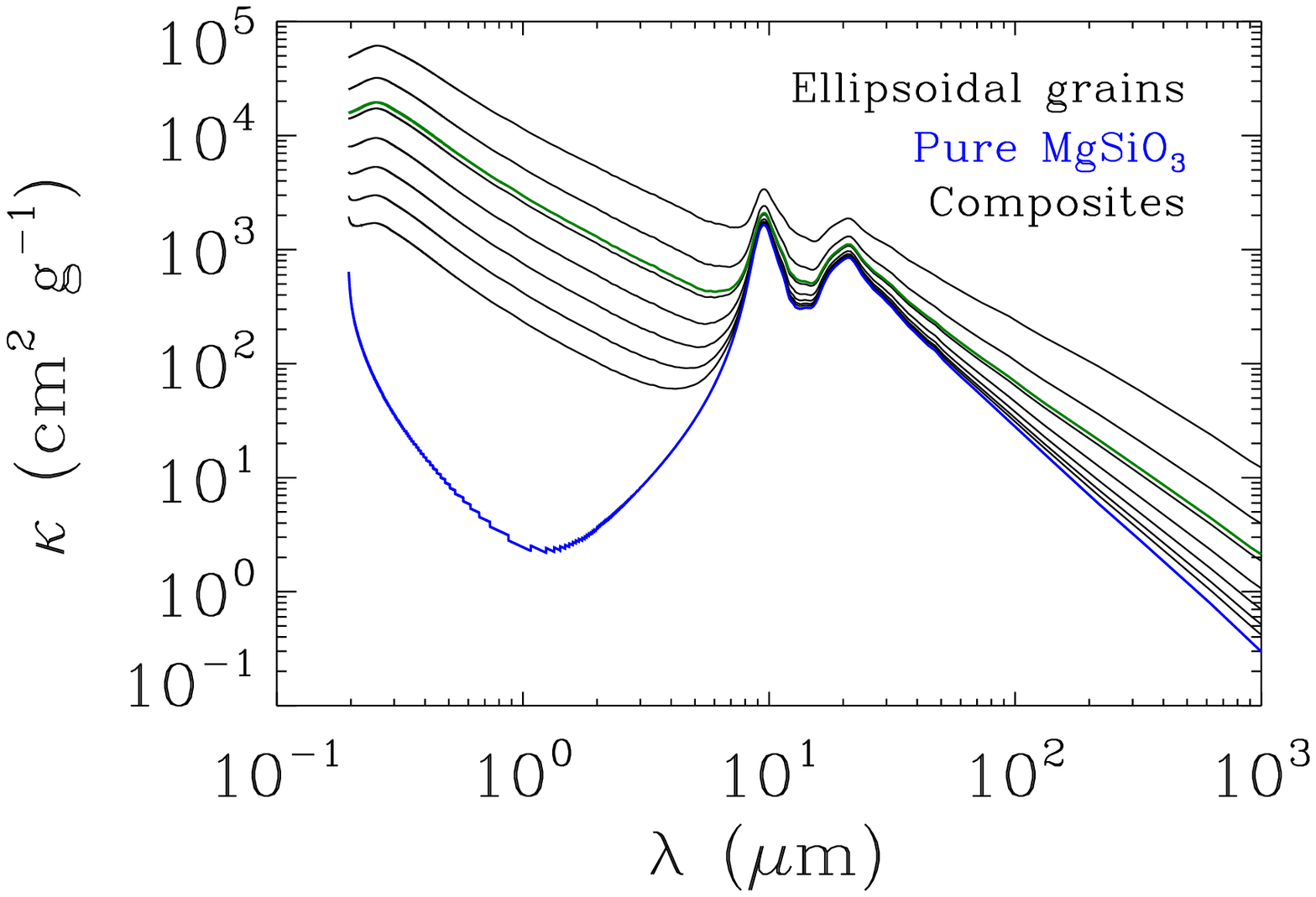} 
   \caption{{\footnotesize The mass absorption coefficients used in this paper. {\bf Left panel}: Spherical silicate (\sild) and amorphous carbon grains; {\bf Right panel}: Ellipsoidal composite grains. The blue curve corresponds to pure ellipsoidal \sild\ grains. Successive curves represent the effect of added AC inclusions with fractional volume filling factor of 0.016, 0.028, 0.05, 0.09, 0.16, 0.28, and 0.50. The green curve represents the $\kappa$ used to fit the spectra of epochs 8515 and 9090, and corresponds to grains with \sild\ and AC volume filling factors of, respectively, 82 and 18\%.}}
\label{kappa}
\end{figure*}  

%================================================================  
  \section{THE EVOLUTION OF DUST MASS}
%================================================================ 

%-----------------------------------
\subsection{Nucleosynthetic Constraints on the Dust Mass}
%-----------------------------------
The mass of dust that can form in the SN ejecta is limited by the mass of refractory elements produced during the quiescent and explosive phases of the evolution of the progenitor star. SN1987A is believed to be the end stage of the evolution of a $\sim 20$~\msun\ star, identified as the blue supergiant Sk~--69~202 \citep{gilmozzi87}. Its initial metallicity is presumably that of the LMC, which is $\sim 40$\% solar or Z=0.008.
Table~\ref{yields} presents the post explosion elemental yields of the major refractory elements in the ejecta. The yields from \cite{nomoto06} were interpolated between the tabulated values for Z=0.004 and 0.02. The yields of \cite{woosley88b} were taken from his initial model with LMC abundances, that was calculated to fit the early SN light curve. Both models predict that Mg and Si are produced in about equal amounts of $\sim 0.1$~\msun, but differ in the mass of predicted carbon, which varies between 0.144 and 0.25~\msun. Both models produce the right amount of Fe (synthesized as $^{56}$Ni) to power the SN light curve. The iron resides in a hot low density radioactive bubble, and is not likely to condense into dust at early times.  We therefore only considered silicate and carbon dust in our models. Since the masses of Mg and Si are about equal, the simplest way to incorporate them in the dust is to assume that the dust consists of SiO$_2$ and MgO clusters or \sild\ grains. In our models we assume that the silicates are in the form of \sild, and the carbon in the form of amorphous carbon (AC) dust.    

%=========================
% Table 1
%=========================
\begin{deluxetable}{lccccc}
\tablewidth{0pt}
%\tabletypesize{\scriptsize}
\tablewidth{0pt}
\tablecaption{Elemental Yields of a 20~\msun\ Star}
\tablehead{
\colhead{Metallicity } &
\colhead{C } &
\colhead{Mg } &
\colhead{Si } &
\colhead{Fe } &
\colhead{Reference\tablenotemark{1}} 
}
\startdata
0.004 & 0.097 & 0.101 & 0.124 & 0.074 & [1] \\
0.020 & 0.245 & 0.095 & 0.068 & 0.089 &  \\
0.008 \tablenotemark{2} & 0.144 & 0.098 & 0.096 & 0.080 &  \\
\hline
0.008  & 0.250 & 0.096 & 0.11 & 0.075 & [2]
\enddata
\tablenotetext{1}{References: [1]-\cite{nomoto06}; [2]- \cite{woosley88b}}
\tablenotetext{2}{Yields for this metallicity were obtained by logarithmic interpolation between Z=0.004 and 0.02}
\label{yields}
\end{deluxetable}

%-----------------------------------
\subsection{Abundances Constrained Fits to the Infrared Spectra of SN1987A}
%-----------------------------------

The inner ejecta of the SN was well resolved with the Atacama Large Millimeter/Submillimeter Array (ALMA) at far-IR wavelengths, and found to have an elliptical shape with dimensions corresponding to freely expansion velocities of $1350\pm 150$ and $750 \pm 250$~\kms, along the major and minor axes of the ellipse \citep{indebetouw14}. For simplicity we will adopt a spherical shape for the ejecta with a constant expansion velocity of $v = (1350\times750^2)^{1/3} = 910$~\kms. The radius of the ejecta at the different epochs is given in Table~\ref{results}. On day 9000 it is about $7\times10^{16}$~cm, and its radial 200~\mic\ optical depth is $\sim 0.3$. 
Since the ejecta is optically thin at these epochs, we start our analysis with the spectra of those epochs.

%-----------------------------------
\subsubsection{Days 8515 and 9090}
%-----------------------------------
 High resolution ALMA images at 450 and 850~\mic\ clearly resolved the emission from the ejecta and the circumstellar ring \citep{indebetouw14, zanardo14b}. Spectral observations, summarized by \cite{matsuura15}, showed that the far-IR/submm fluxes obtained by {\it Spitzer} and ALMA observations are dominated by continuum emission from ejecta dust, except for the 70~\mic\ flux, which may be contaminated by the [O~I]~63\mic\ line and by the continuum emission from the hot dust in the circumstellar ring \citep{dwek10}.  The combined {\it Spitzer} and ALMA measurements of the far-IR and submillimeter  flux densities of SN1987A used in our analysis were taken from  \cite[][their Table 1]{matsuura15}. We adopt here their estimate that at least 2/3 of the observed flux in the 70~\mic\ band arises from the ejecta dust.

To minimize the mass of dust in the ejecta, \cite{matsuura15} considered the possibility that the dust grains are in the shape of elongated ellipsoids. Elongated grains radiate more efficiently at long wavelengths, thereby lowering the mass needed to produce the emission. However, the shape effect in AC grains does not only increase their long wavelength emissivity, it also considerably flattens it with wavelength. Since most of the emission in the Matsuura model is attributed to AC dust, the resulting spectrum became too broad to fit the observations.
To circumvent these problems we adopt a dust model consisting of ellipsoidal composite grains, comprising mostly of a silicate (\sild) matrix with AC inclusions.
The elongated shape increased the dust emissivity, and the addition of AC inclusions further increased the dust emissivity, with only minor flattening of its wavelength dependence.

Figure~\ref{kappa} depicts the various mass absorption coefficients considered in our models. In the Rayleigh limit, where $a << \lambda$, the mass absorption coefficient of the dust is independent of grain radius. 
The left panel shows the wavelength dependence of $\kappa(\lambda)$ for pure spherical \sild\ (blue curve) and amorphous carbon (red curve) grains. Optical constants for the silicate dust were taken from \cite{jager03}, and those for the AC dust from \citep{zubko96}. The right panel depicts the wavelength dependence of the ellipsoidal composite grains. The blue curve corresponds to pure \sild\ grains. The black curves give the values of $\kappa$ for the composite grains with different amount of AC inclusions (see figure caption). The green curve corresponds to the mass absorption coefficient of the composite grains used to fit the 8515 and 9090 epochs spectra.  

\begin{figure}[tbp]
  \centering
   \includegraphics[width=3.0in]{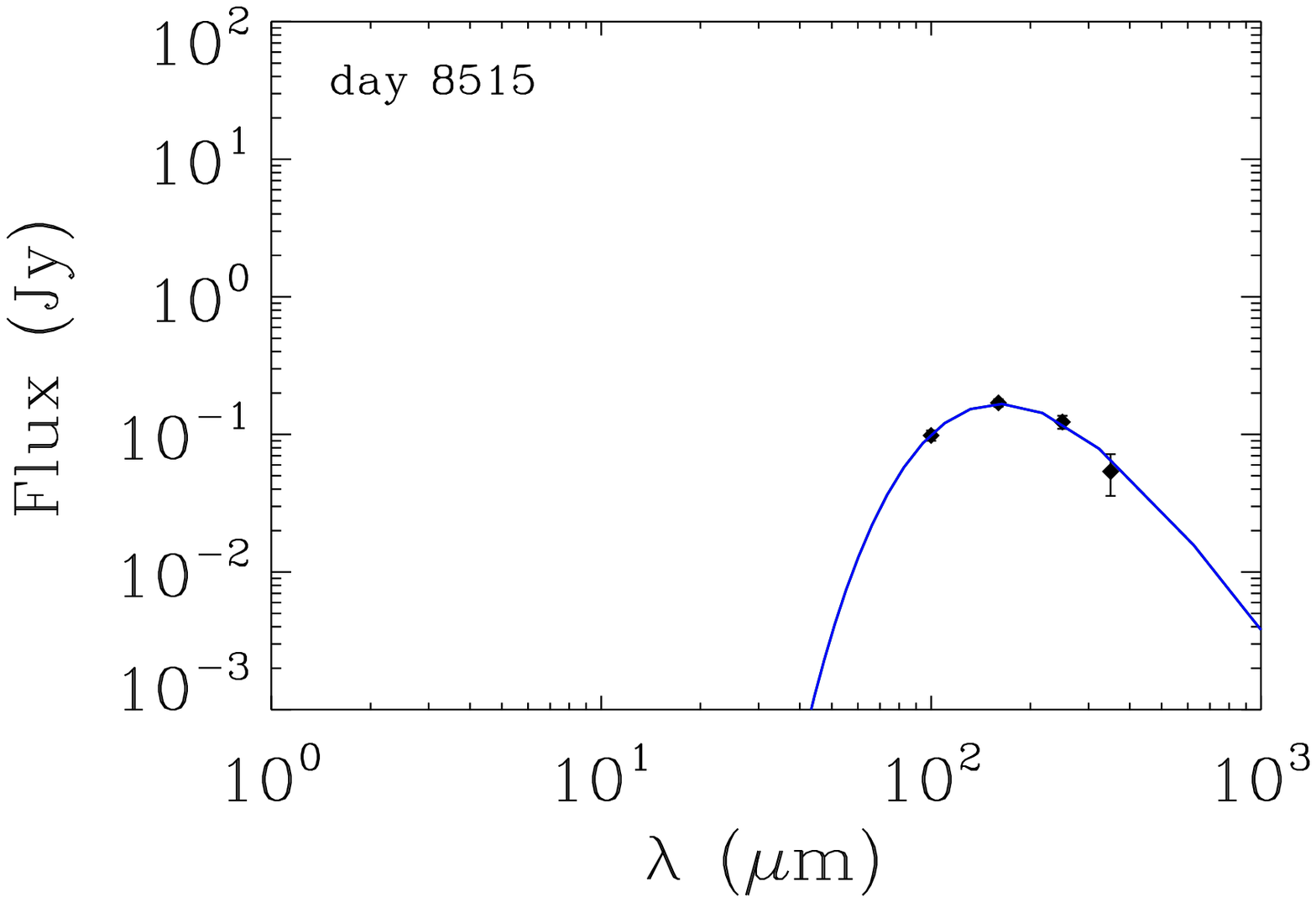}
      \includegraphics[width=3.0in]{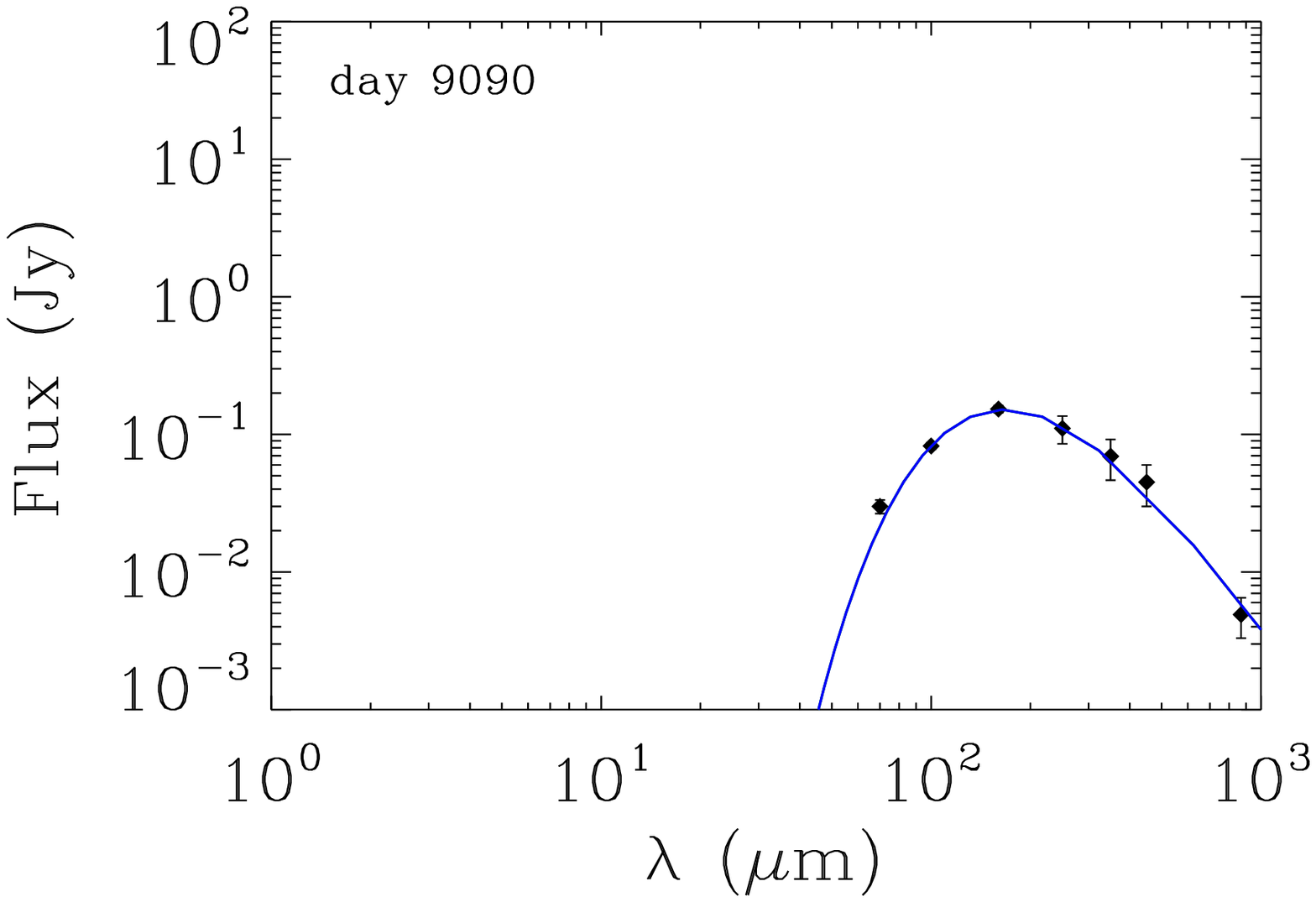}\\
 \caption{{\footnotesize  Fits to the observed spectrum with composite grains consisting of MgSiO$_3$ and AC with volume filling factors of 82 and 18\%, respectively. Detailed description of the fits and tabulated masses are given in the text and in Table~\ref{results}. }}
\label{spec1}
\end{figure}

We assumed that the dust radiates at a single temperature and used the IDL routine MPFIT \citep{markwardt09} to simultaneously fit the SN1987A flux densities for the two epochs, with the dust masses and temperatures as variables. We experimented with different concentrations of AC inclusions. The results presented here are for composite grains with \sild\ and AC volume filling factors of $f_s=0.82$ and $f_c=0.18$, respectively. The choice of parameters was driven by the constraints on the available mass of Mg and Si in the ejecta. The resulting spectra are shown in Figure~\ref{spec1}, and the resulting dust masses and temperatures are summarized in Table~\ref{results}. 

The mass fraction of \sild\ and AC in the grain are given by $f_s\, (\rho_s/\overline{\rho})$ and $f_c\, (\rho_c/\overline{\rho})$, respectively, where $\rho_s=3.2$~g~cm$^{_3}$ and $\rho_c=1.8$~g~cm$^{_3}$ are, respectively, the mass densities of \sild\ and AC grains, and $\overline{\rho}=f_s\rho_s+f_c\rho_c$ is the average density of the grain.
If $M_d$ is the total dust mass, then the mass of Mg, Si, and C locked up in the dust is given by $0.24\,f_s\, (\rho_s/\overline{\rho})\, M_d$, $0.28\,f_s\, (\rho_s/\overline{\rho})\, M_d$, and $f_c\, (\rho_c/\overline{\rho})\, M_d$, respectively. For an AC filling factor of 0.18 we get $\overline{\rho}=2.95$~g~cm$^{_3}$, \sild\ and AC mass fractions of 0.89 and 0.11, respectively, and Mg and Si mass fractions of 0.21 and 0.25, respectively. The elemental masses locked up in the dust are presented in Table~\ref{results}. About half of the \sild\ mass consists of O which, because of its large abundance in the ejecta ($\sim 1$~\msun), does not impose any constraints on the mass of dust that can form in the ejecta.   

The results show that the late-time spectra require $\sim 0.09$~\msun\ of Mg, $\sim 0.1$~\msun\ of Si, and $\sim 0.046$~\msun\ of C to be locked up in dust. From Table~\ref{yields}, we see that these masses are well within their expected yields, alleviating the abundance problems of previous models.    

%------ figure 2

%=========================
% Table 2
%=========================
\begin{deluxetable*}{lccccc}
\tablewidth{0pt}
%\tabletypesize{\scriptsize}
\tablewidth{0pt}
\tablecaption{Evolution of Dust Parameters and Ejecta Radius\tablenotemark{1}}
\tablehead{
\colhead{Epoch (day)} &
\colhead{615} &
\colhead{775} &
\colhead{1144} &
\colhead{8815} & 
\colhead{9090}   
}
\startdata
Pure silicates (MgSiO$_3$)          & & & & \\
\hline
M$_d$(\msun)                   &   {\bf 0.40}     &  {\bf 0.40}    &  {\bf 0.40} &  \nodata   &  \nodata    \\
T$_d$(K)                       &   607      &   334     &   145      & \nodata    &  \nodata \\
\hline  
Pure amorphous carbon          & & & & \\
\hline
M$_d$(\msun)                      &   {\bf 0.047}   &  {\bf 0.047}    &   {\bf 0.047}      & \nodata    &  \nodata   \\
T$_d$(K)                       &   454      &  333      &   {\bf 120}  &  \nodata   &  \nodata  \\
\hline
Composite grains \tablenotemark{2}         & & & & \\
\hline
M$_d$(\msun)                      &  \nodata  &  \nodata   &   \nodata  & $0.42\pm 0.07$    &  $0.45\pm 0.08$  \\
T$_d$(K)                       &  \nodata  &  \nodata   &   \nodata  &  26.3   &  25.2  \\
\hline
Elemental abundances in dust         & & & & \\
\hline
Mg(\msun)                            &   0.096     &  0.096     &   0.096  &  0.090    &  0.095   \\
Si(\msun)                            &   0.11      &  0.11      &   0.11   &  0.11     &  0.11   \\
C(\msun)                             &   0.047     &  0.047     &   0.047  &  0.045    & 0.048   \\
\hline
%-------------------------------------------------------------------------------------------------------
\hline
R$_{ej}$ [$10^{16}$ cm]\tablenotemark{3}     &  0.48             & 0.61              &  0.90             & 6.9          & 7.1     \\
L$_{IR}$ (\lsun) \tablenotemark{4}           &  $3.6\times10^5$  &  $1.2\times10^5$  & $7.0\times10^3$   &  310$\pm$27  &  270$\pm$19  \\
$\tau(\lambda)$\tablenotemark{5}             &   9130            &  5650              &   252            &  0.33        &  0.33   
\enddata
 \tablenotetext{1}{Entries in bold were held constant. The spectra from the last two epochs were fit with ellipsoidal composite grains. Grains in the first three epochs consist of a mixture of spherical silicates (MgSiO$_3$), and amorphous carbon (AC) grains.} 
 \tablenotetext{2}{The composite grains consist of an MgSiO$_3$ matrix with AC inclusions occupying 18\% of the grain's volume.}
\tablenotetext{3}{The ejecta radius, R$_{ej}$, at each epoch was calculated for a constant expansion velocity of 910~\kms.}
\tablenotetext{4}{For comparison, on day 1144 the U to 20~\mic\ (uvoir) luminosity is 3400~\lsun\ \citep{suntzeff91}.}
\tablenotetext{5}{The optical depth, $\tau(\lambda)$ is given by Equation~(\ref{tau}), and calculated at around the peak wavelength of the emission for each epoch, at 20, 20, 50, 200, and 200~\mic, respectively.}
\label{results}
\end{deluxetable*}

%-----------------------------------
\subsubsection{Days 615, and 775}
%-----------------------------------
The most detailed spectra of SN1987A during its early evolutionary stages were presented by \cite{wooden93}. As mentioned before, the lack of spectral feature has been construed as evidence for the lack of silicate dust in the ejecta. To test this conclusions we ran models with pure spherical silicate (\sild) and AC grains, in which the mass of Mg and Si in the dust was taken to be equal to that derived for epochs 8515 and 9090. For this choice of dust masses, or any other sufficiently large dust mass, the ejecta is optically thick, and consequently the shape of the dust grains will not affect the blackbody spectrum of the ejecta. Our choice of using distinct silicates and AC dust populations is motivated by the dust growth and coagulation models of \citep{sarangi15}. In these models dust grows primarily by accretion at early epochs, after which the abundance of condensible elements is sufficiently depleted and coagulation becomes the main process for grain growth.
      
\begin{figure}[tbp]
  \centering
            \includegraphics[width=3.0in]{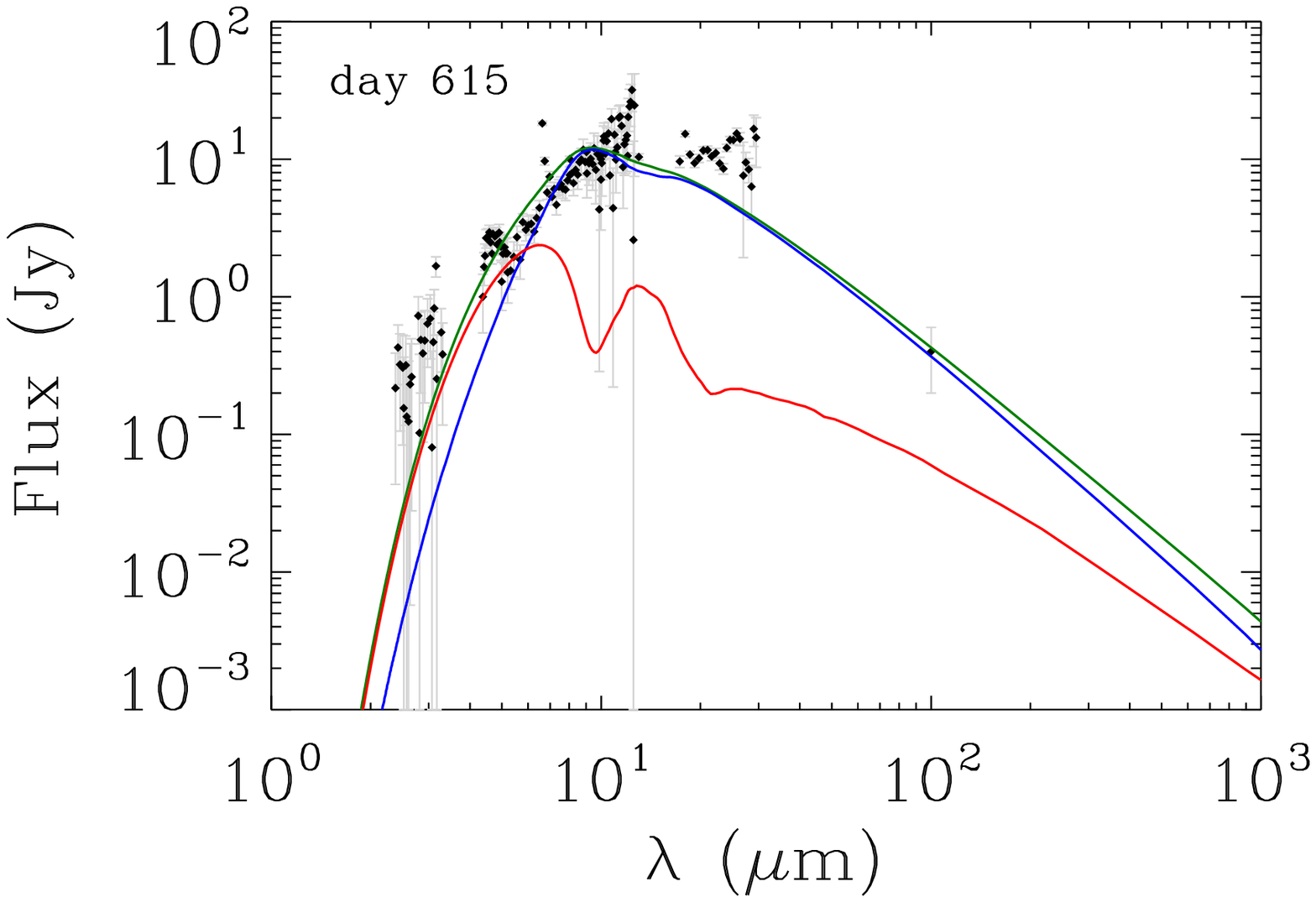} 
               \includegraphics[width=3.0in]{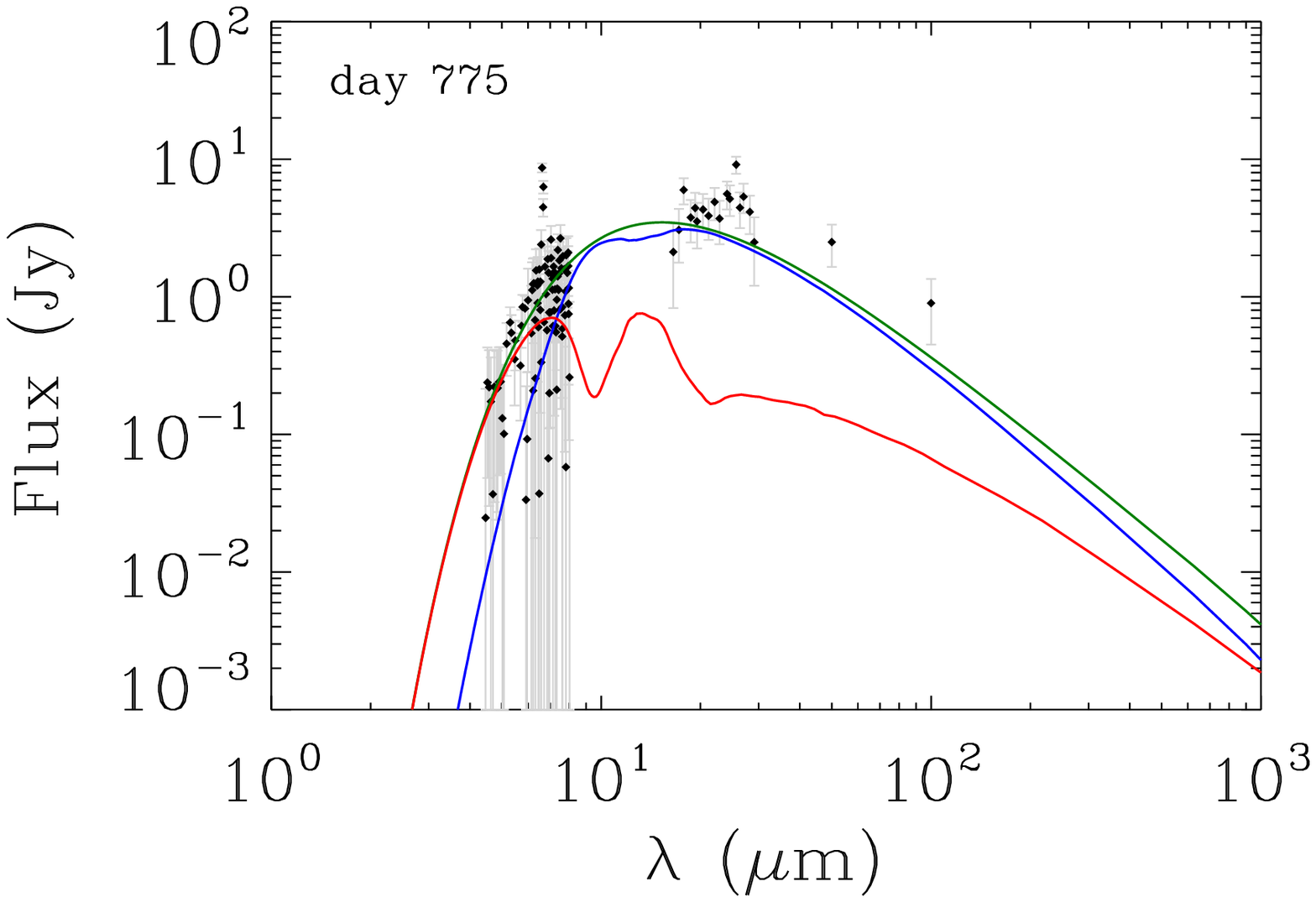}\\
 \caption{{\footnotesize  The spectra on days 615 and 775, obtained with a combination of pure MgSiO$_3$ silicates (blue line) and amorphous carbon (AC) grains (red line). Masses of the two components were fixed and determined by the masses of Mg, Si, and C locked up in dust in epochs 8515 and 9090. The total flux is given by the green line. Detailed description of the fits and tabulated masses are given in the text and in Table~\ref{results}. }}
\label{spec2}
\end{figure}

Thus the only free parameters were the silicate and AC dust temperatures. The resulting spectra are shown in Figure~\ref{spec2}. 
The red and blue curves are, respectively, the  carbon and silicate emission components of equation~(\ref{flux2}), but only their sum (green curve) is observable.  
The 20~\mic\ optical depth of the ejecta is $\sim 10^4$ and 6000 for these two epochs (see also Table~\ref{results}), giving rise to  smooth featureless spectra. 
They provide a surprisingly good fit to the observations considering the fact that we forced each of the silicate and carbon dust to radiate at single (but different) temperatures.  The figure shows that the silicate 9.7 and 18~\mic\ emission features have been internally absorbed by the silicate-carbon dust mixture, as evident from the strong absorption features in the carbon dust emission.

%------ figure 3

%-----------------------------------
\subsubsection{Day 1144}
\begin{figure}[tbp]
  \centering
         \includegraphics[width=3.0in]{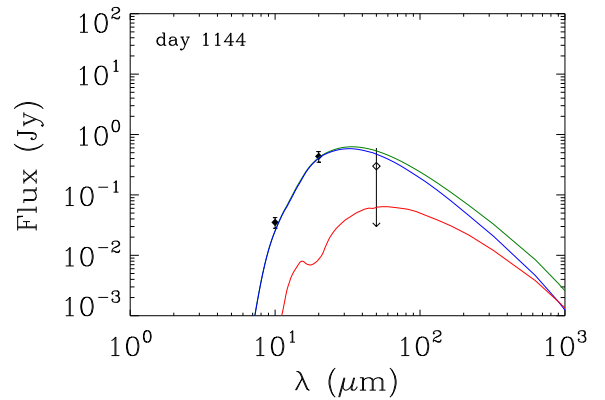}\\
 \caption{{\footnotesize  The spectrum corresponding to the same dust compositions as in epochs 615 and 775, but with temperatures constrained to obey the observed upper limit on the IR flux at 50~\mic. Detailed description of the fits and tabulated masses are given in the text and in Table~\ref{results}. \label{spec3}}}
\end{figure}

%-----------------------------------
We selected the 10 and 20~\mic\ photometric data obtained with the Cerro Tololo Inter-American Observatory (CTIO) on day 1144 to represent the observations taken during the 935 to 1352~d interval \citep{suntzeff91}. For day 1144 we also have a 50~\mic\ upper limit reported by P. M. Harvey, and presented by \cite{dwek92c}.  A low resolution ($\lambda/\Delta \lambda = 40$) 16-30~\mic\ spectrum of the SN was obtained with the KAO around day 1151 \citep{dwek92c}. Several emission lines were detected, and the upper limit set on the continuum intensity is consistent with the CTIO 20~\mic\ detection. Figure~\ref{spec3}  presents the CTIO data and the 50~\mic\ upper limit. As for days 615 and 775, we assumed that the dust giving rise to the emission consists of a combination of spherical \sild\ and AC grains with masses identical to this in epochs 615 and 775, and allowed the temperature to vary. Figure~\ref{spec3} shows the fit to the observed spectrum. The dust masses and temperatures are presented in Table~\ref{results}.  

% 

%------ figure 4

%----------------------------------------------------------------
\section{Clumpy Ejecta}
%----------------------------------------------------------------
So far we assumed that the ejecta can be approximated by a dusty sphere of radius $R$. However, the early detection of X- and $\gamma$-rays from the SN, and the high ionization rate in its envelope, required the macroscopic mixing of clumps containing radioactive Ni, Co, and Fe throughout the ejecta \citep{graham88,li93}. Furthermore, the lack of silicate emission features in the early IR spectra (days 615, 775) was attributed to the fact that the dust may have formed in optically thick clumps \citep{lucy89, lucy91}. 
In the following we show that the results of our calculations can be simply extended to a clumpy medium.

In the following we will construct a simple model for a clumpy ejecta. Since the ejecta is optically thin at late epochs (days 8515 and 9090), the model is relevant only for the epochs when the homogeneous ejecta was optically thick.  Using the same dust mass and composition as in the homogeneous model, we show that the clumpy model can reproduce the same IR spectrum at each epoch, and is also capable of accounting for the UVO region of the SN light curve. 

We will assume the presence of $N_c$ identical dusty clumps with a dust mass $m_c=M_d/N_c$, where $M_d$ is the total dust mass in the ejecta, and radius $r_c$. We also assume that the clumps maintain pressure equilibrium with their surroundings so that the ratio $r_c/R$, where $R$ is the outer radius of the ejecta, remains constant with time. We also assume the presence of some dust mass, $m_{ic}$, in the interclump medium. 

The volume filling factor of the clumps is given by:
\begin{equation}
\label{fvol}
f_V = N_c\, x^3 \qquad . 
\end{equation}

The number of clumps and their radii are constrained by the requirement that the projected area of the clumps be equal to that of the  the homogeneous sphere.  This requirement reduces to the constraint that $N_c\, r_c^2 = \xi\, R^2$, where $\xi$ is a factor of order unity that compensates for the overlapping of clumps along a given line of sight. 

 The optical depth of each clump, $\tau_c$, is simply related to $\tau_h$, that of the homogeneous sphere, by
\begin{equation}
\label{tauc}
\tau_c = {3\over4}\, \left({M_d/N_c\over \pi r_c^2}\right)\,\kappa = {\tau_h\over \xi} 
\end{equation}
The clumps are therefore optically thick at UVOIR wavelengths at all early epochs before day 1144, so that their cross section can be taken as equal to the geometrical value $\pi\, r_c^2$.

The probability that at photon passes a distance $z$ through the ejecta without absorption, and then hits a clump in the $z$ and $z+dz$ interval is given by:
%-------------------
\begin{equation}
\label{px}
P(z)\, dz = \exp(-z/\ell)\, {dz \over \ell}
\end{equation}
%-------------------
where $\ell = (n_c \pi r_c^2)^{-1}$ is the photon mean free path through the ejecta, and $n_c$ is the number density of clumps.
A photon incident on the sphere within an annulus $2\pi\, b\, db$, where $b$ is the projected distance from the center of the sphere, will strike a clump with a probability given by:
%-------------------
\begin{eqnarray}
\label{pb}
P(b)\ db & =  & \int_0^{2R\sqrt{1-(b/R)^2}}\ \exp(-z/\ell)\, {dz \over \ell} \\ \nonumber
  & = & 1-exp\left(-2R\, \sqrt{1-(b/R)^2}\right)
\end{eqnarray}
%-------------------
The fraction $f_A$ of the projected area of the ejecta that is filled by the clumps is then given by the integral:
\begin{equation}
\label{fa}
f_A = (\pi R^2)^{-1}\int_0^R\ P(b)\, 2\pi\, b\, db 
\end{equation}

\begin{figure}[tbp]
  \centering
         \includegraphics[width=3.0in]{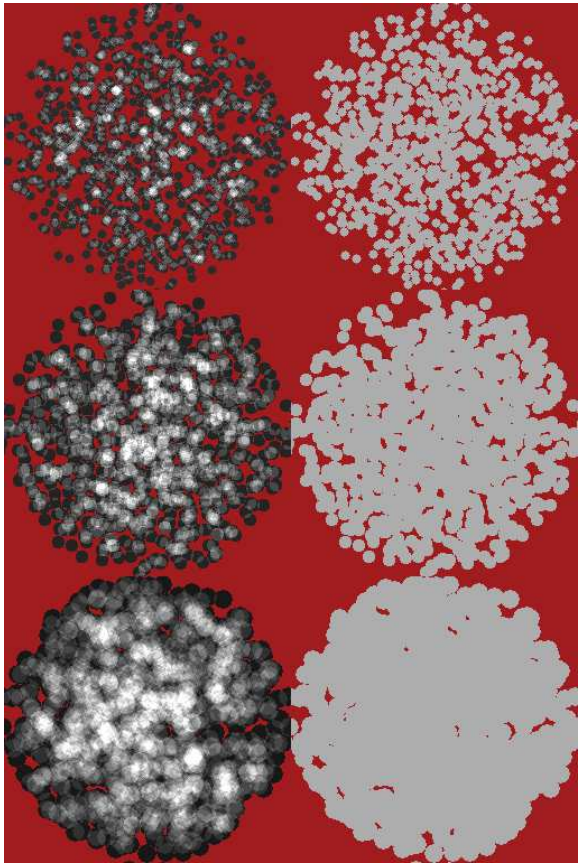}\\
 \caption{{\footnotesize  Monte Carlo simulations of clumpy ejecta characterized by $N_c=1000$ and $\xi$=1.0, 2.0, and 4.0 (top to bottom). The column on the left depicts the dust column density. The brighter regions depict regions of overlapping clumps. The right column depicts the projected surface area of all the clumps. The surface covering factors, $f_A$ for this simulation are listed in Table~3.}\label{clumps2}}
\end{figure}

Table~\ref{clumps} summarizes some clump characteristics and the fraction of the projected surface of the ejecta covered by clumps for several values of $N_c$ and $\xi$. The most desirable combination of these parameters is one that gives a small volume filling factor for the clumps, and a large surface covering factor. The combination of \{$N_c, \xi$\}=\{1000, 2\} fills both requirements. It give a clump filling factor of 0.09, a surface covering fraction of 0.82. The value of $r_c/R$ for this model is 0.044, consistent with the scales of clumps predicted by numerical simulations of the instabilities in core collapse supernovae \citep{wongwathanarat15}. This means that in the clumpy model, the integrated brightness of the IR emission will be lower by 18\% compared to that of the homogeneous sphere. This factor is well within the uncertainties in the fluxes observed on days 615-1144. So all dust masses derived assuming  homogeneous ejecta are still valid for the clumpy ejecta model. Figure~\ref{clumps2} presents Monte Carlo simulations of a clumpy ejecta for $N_c=1000$ clumps and a values of $\xi$=1.0, 2.0, and 4.0 (top to bottom). The simulations were performed to validate the analytically derived area covering fractions listed in Table~\ref{clumps}. From top to bottom, the fraction of the projected area covered by the clumps is 0.59, 0.82, and 0.96, in  good agreement with the analytical formula presented in Equation~(\ref{fa}).

%------ figure 5

 The clumpy model must allow for the escape of UVO photons, and any selective extinction in the ejecta. In the following we use the mega-grain approximation described by \cite{varosi99} to calculate the escape of these photons from the clumpy medium. 
Equation~(59) in their paper can be simplified by assuming that the volume filling factor, $f_V$, of the clumps is small, that most of the dust resides in the clumps, and that the clumps are solid spheres with geometrical cross section $\pi r_c^2$. The radial optical depth, $\tau_R(\lambda)$, of the ejecta is then given by the sum of two terms. The first is wavelength independent and caused by the absorption in the optically thick clumps, and the second is the wavelength dependent absorption by  dust residing in the interclump medium.
\begin{eqnarray}
\label{taur}
\tau_R(\lambda)&  = & n_c\, \pi r_c^2\, R + \rho_{ic}\, \kappa(\lambda)\, R   \\ \nonumber
& = & (3/4) \left[ \xi +  (m_{ic}/\pi R^2) \, \kappa(\lambda) \right]
\end{eqnarray}
where $\rho_{ic}$ and $m_{ic}$ are, respectively, the density and mass of the dust in the interclump region, and $V$ is the volume of the ejecta.

At early times, the IR emission accounts for half the energy deposited in the ejecta by the radioactive decay of $^{56}$Co. So the unabsorbed UVO luminosity can be at most equal to that of the IR. The observed UVO flux constitutes about 5-10\% of the IR emission from the SN \citep[][Figures 1 and 2]{lucy89,lucy91,wesson15}. So at least 5 to 10\% of the unabsorbed UVO emission must escape the interclump medium. These, and observations of nebular emission lines from the ejecta \citep{jerkstrand11}, constrain the mass of dust that can be present in this medium. 

For radial optical depths $\gtrsim 1$, the escape probability from the ejecta is approximately given by $3/4\tau$. For $\xi = 2$, the first term in Equation~(\ref{taur}) contributes $3\xi/4 = 1.5$ to $\tau_R$, giving $P_{esc} \approx 0.5$. For $P_{esc}$ to be larger than 0.05, the optical depth must be smaller than $\sim 15$, setting a limit of  $ \tau \lesssim  13$ on the second term in eg.~(\ref{taur}).
The mass absorption coefficient of the dust at 0.55~\mic\ is about $10^4$~cm$^2$~g$^{-1}$ for a mixture of silicate-carbon grains (Figure 1), so the second term becomes 
\begin{equation}
\label{ }
(3/4)\, (m_{ic}/\pi R^2) \, \kappa(V) \approx 2\times10^5\, m_{ic}({\rm M}_{\odot}) \lesssim 13
\end{equation} 
where we assumed an ejecta radius of $0.5\times10^{16}$~cm. So the interclump dust mass is limited to $m_{ic}\lesssim 10^{-4}$~\msun, allowing  UVO emission that is not absorbed in the clumps to escape the ejecta. 

%=========================
% Table 3
%=========================
\begin{deluxetable}{ll|lll}
\tablewidth{0pt}
%\tabletypesize{\scriptsize}
\tablewidth{0pt}
\tablecaption{Surface covering factor \\for several clump models}
\tablehead{
\colhead{$N_c$} &
\colhead{$\xi$} &
\colhead{$f_V$} &
\colhead{$r_c/R$} &
\colhead{$f_A$} 
}
\startdata
100   & 1   & 0.10 & 0.10 & 0.60  \\
      & 2   & 0.28 & 0.14 & 0.82  \\
      & 4   & 0.80 & 0.20 & 0.94  \\
      \hline
1000   & 1   & 0.03 & 0.03 & 0.60  \\
      & 2   & 0.09 & 0.044 & 0.82  \\
      & 4   & 0.25 & 0.063 & 0.94  
\enddata
\label{clumps}
\end{deluxetable}

In summary, the clumpy model can reproduce most of the early IR emission from the ejecta with the same mass of dust as in the homogeneous model. The clumps are optically thin at late times, so that the mass derived for days 8515 and 9090 is unaltered as well. 
The mass of dust in the interclump region is small compared to that in the clumps, allowing for the escape of nebular and continuum UVO emission from the ejecta. Because of the small mass of dust in the interclump medium, it makes no significant contribution to the IR emission from the clumps.

%================================================================  
  \section{THE EVOLUTION OF THE IR LUMINOSITY AND THE  $^{44}$Ti ABUNDANCE IN THE EJECTA}
%================================================================ 

%---------------------------------------
\subsection{The Evolution of the IR Luminosity}
%---------------------------------------   

The IR luminosity at each epoch is simply given by the integral of the spectrum, assuming a source distance of of 50~kpc. The results are given in Table~\ref{results}. At early times, the IR luminosity follows the $^{56}$Co decay curve, but is consistent with being constant on days 8515 and 9090.

The energy absorbed and re-radiated by the dust could be either generated by sources external to the ejecta, or by internal energy sources.
External energy sources include the X-ray and UVO emission generated by the reverse shock \citep{france15}, and/or the diffuse interstellar radiation field and nearby stellar sources. Internal energy sources include a potential embedded pulsar or decaying radioactive isotopes.

%=========================
% Table 4
%=========================
\begin{deluxetable}{llll}
\tablewidth{0pt}
\tablecaption{Inferred $^{44}$Ti Yield from UVOIR and FIR Observations of SN1987A}
\tablehead{
\colhead{M($^{44}$Ti) } &
\colhead{Observations} &
\colhead{Epoch} &
\colhead{Reference\tablenotemark{1} } \\
\colhead{ ($10^{-4}$~\msun) } &
\colhead{ } &
\colhead{(days)} &
\colhead{ } 
}
\startdata
$3.1\pm0.8$   & hard X-rays    & 3600-4000    &  [1]   \\
1.5           & UVOIR lines    & $\sim$3000   &  [2] \\
$0.55\pm0.17$ & UVOIR          & $<$1900      &  [3]   \\
              & extrapolated V & $<$4300      &                      \\
1.6           & far-IR dust    & 8500-9100    &  this work 
\enddata
\label{ti44}
\tablenotetext{1}{References: [1] \cite{grebenev12};  [2] \cite{jerkstrand11}; [3] \cite{seitenzahl14}}
\end{deluxetable}

%---------------------------------------
\subsection{External Energy Sources}
%---------------------------------------   

\subsubsection{The reverse shock}
Prediction that the reverse shock generated by the interaction of the SN blast wave with the circumstellar ring would be manifested in increased line emission from the ejecta were confirmed by UVO imaging spectroscopy of SN1987A with the {\it Hubble Space Telescope} ({\it HST}) \citep[][and references therein]{france11}. During the period of interest here ($\sim$ day 9700), the X-ray - UV emission
of the ER and the reverse shock over the  0.01-8 keV energy range is $\sim 1460$~\lsun\ 
\citep[day 9885,][]{france15}, apparently sufficient to power the IR emission. However the solid angle subtended by the ejecta 
($\sim 2 \times 10^{17} \times 1 \times 10^{17}$ cm), as seen from the equatorial ring (ER)
(radius $= 6.1 \times 10^{17}$~cm) is only 0.158 sr. So even if all this incident radiation traverses the dust-free external H-layers and gets absorbed by the dust,  only $\sim1.3\%$ of this luminosity can 
be absorbed and reradiated by the ejecta. 

\subsubsection{Diffuse and discrete stellar sources}
Other external heating of the ejecta can be supplied by the interstellar 
radiation field (IRSF). However, if the energy density is the same as in the 
solar neighborhood \citep{mathis83, galliano11}, then the 
incident flux is 55 times weaker than the 
(unattenuated) X-ray flux. 

Nearby Stars 2 and 3 are B2 III stars with luminosities of 
$\sim 2 \times 10^4$ and $\sim 1\times 10^4\ L_\sun$ 
\citep{walborn93}. If these stars
are at the same line of sight distance as SN 1987A, then the stars can 
provide fluxes as large as 1 and 1.5 times (respectively) 
the flux of the X-rays. 

Altogether, the X-ray emission can at most provide $1460\times 0.158/4\pi = 18$~\lsun. The local ISRF can provide only 0.3~\lsun, and stars~2 and 3, at most 18 and 27~\lsun, respectively. 
So at most, $\sim 63$~\lsun\ of the total IR emission of $\sim 300$~\lsun, can be supplied by external energy sources. 
  
%---------------------------------------
\subsection{Internal Energy Sources}
%---------------------------------------   

\subsubsection{Compact central object}
One internal source of energy would be from accretion onto or 
spin down of a pulsar. Searches for a compact remnant have been unsuccessful (except possibly Middleditch et al. 2000), and
have set low luminosity limits on a compact source 
($\lesssim 2L_\sun$) \citep{graves05}. However, this
could be higher if the extinction is far larger than was assumed.

\subsubsection{Decaying radioactive elements} 
During the first two years after the explosion, the SN light curve was powered by the energy released in the radioactive decay chain $^{56}$Ni$\longrightarrow ^{56}$Co$\longrightarrow ^{56}$Fe. Models show that  $\sim 0.07$~\msun\ of $^{56}$Ni were required to fit the early UVOIR light curve, before dust has formed in the ejecta \citep[][and references therein]{woosley88b, seitenzahl14}. After about 2000 days the SN1987A light curve is predominantly powered by the decay of $^{44}$Ti ($\tau_{1/2}=58.9$~yr). 
Efforts to determine the amount of $^{44}$Ti required to power the late time light curve have focused on hard X-rays, nebular lines, and UVOIR observations of the SN. A detailed summary is given by \cite{seitenzahl14}, with highlights given in Table~\ref{ti44} of this paper.
Until day $\sim 1800$ the luminosity was determined from the UVOIR observations, which extend out to the 20~\mic\ band.  
However, most of the dust emission after day $~\sim 1100$ is emitted at longer wavelengths (see Figure~\ref{spec3}, so that he UVOIR luminosity may be missing a significant fraction of the total energy emitted by the ejecta. 

\begin{figure}[tbp]
  \centering
   \includegraphics[width=3.0in]{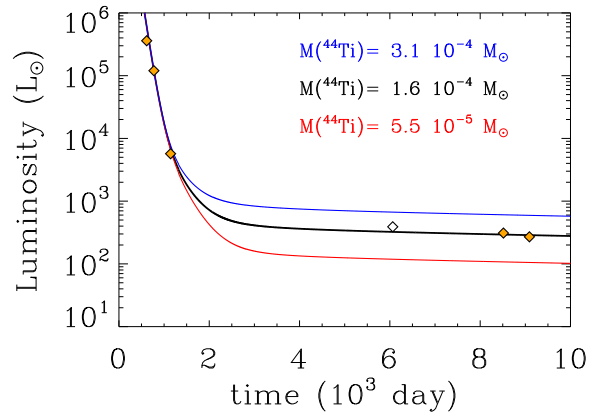}
 \caption{{\footnotesize  The energy released by the decay of radioactive elements in the ejecta as a function of time. All luminosities were calculated for a $^{56}$Ni abundance of 0.071~\msun, but with different $^{44}$Ti abundances, spanning the range of values in Table~\ref{ti44}. The yellow diamonds represent the IR luminosities emitted by the dust at the different epochs (Table~\ref{results}). The black curve represents the best fit to the luminosities on days 8515 and 9090. The open diamond represents the T-ReCS observations on day 6067 \citep{bouchet04}. \label{lum}}}

\end{figure}

Figure~\ref{lum} compares model light curves to the IR luminosities (yellow diamonds). The $^{56}$Ni, $^{57}$Ni, $^{55}$Co, and $^{60}$Co, and masses were taken to be equal to 0.071, $4.1\times10^{-3}$, $9.2\times10^{-6}$, and $4.5\times10^{-8}$~\msun, respectively \citep{seitenzahl14}. 
The figure shows that at early epochs (days 615, 775), the IR emission accounts for about half the radioactive energy deposited by the decay of $^{56}$Co in the ejecta. At later epochs, the IR luminosity can provide an important measure of the amount of $^{44}$Ti created in the explosion.
The different curves correspond to different masses of $^{44}$Ti with values of $0.55\times10^{-4}$ \citep[][red curve]{seitenzahl14}, and $3.1\times10^{-4}$ \citep[][blue curve]{grebenev12}.
Also shown in the figure is the luminosity of 390~\lsun\ on day 6067, derived from the T-ReCS observations of the SN ejecta \citep[][open diamond]{bouchet04}. The black curve provides the best fit to the IR luminosities derived in this paper and requires $M(^{44}$Ti)$=1.6\times10^{-4}$~\msun. 

%------ figure 6
%-------------

This value is in good agreement with the $1.5\times10^{-4}$~\msun\ derived by \cite{jerkstrand11} from the analysis of UVO emission lines in the ejecta around day 3000.  
However, in their standard model they adopted a dust optical depth of only 1. The IR optical depth on that day could be around 10 at the peak of the IR emission, and significantly higher at UVO wavelengths. The $^{44}$Ti mass inferred from the energy required to power the nebular emission lines should therefore be increased to account for the effect of dust absorption. 

On the other hand, the $^{44}$Ti mass inferred from the IR emission alone should also be modified: increased to account for any UVO emission, and decreased to account for the contribution of external heating sources to the emission. The mass should be definitely increased by $\sim 2$\%, to account for the UVO luminosity of $\sim 7$~\lsun\ observed on day 8328 \citep[][Figure 5]{larsson13}, and at most decreased by $\sim 20$\% to account for the possible contribution from external heating sources. So the IR observations provide a robust lower limit of $\sim 1.3 \times10^{-4}$~\msun\ on the mass of $^{44}$Ti synthesized in the explosion.

%================================================================  
  \section{SUMMARY AND ASTROPHYSICAL IMPLICATIONS}
%================================================================ 
In this paper we presented a new scenario for the evolution of dust in SN1987A.
In this scenario, the observed IR spectrum on days 8515 and 9090 arises from $\sim 0.45$~\msun\ of composite ellipsoidal grains, consisting of a mixture of \sild\ and amorphous carbon with volume filling factors of 82 and 18\%, respectively.
In contrast with previous models \citep{matsuura15}, this model requires only $\sim 0.1$~\msun\ of Mg and a similar mass of Si to be locked up in dust, consistent with the post-explosive mass of these elements in the SN ejecta. The mass of carbon locked up in dust is only 0.04~\msun, in contrast to the 0.5~\msun\ of C in the \cite{matsuura15} models.

We also show that the same mass of Mg, Si, and C could have already been locked up in the form of spherical \sild\ and AC grains since day 615 in IR-optically thick clumps. The low mass of dust and the upper limit on the silicate dust in previous models were inferred from the absence of the 9.7 and 18~\mic\ silicate emission features, assuming optically thin clumps. However, we show that the absence of the silicate absorption features at the early epochs is the result of the large optical depth and self-absorption in the clumps. Figures~\ref{spec1}, \ref{spec2}, and \ref{spec3} and Table~\ref{results} show, respectively, the fit to the observed spectra, and the dust parameters corresponding to the fits. 

The model presented in this paper thus solves three problems that plague previous models \citep[][and refernces therein]{matsuura15,wesson15}:
\begin{enumerate}[noitemsep,nolistsep]
  \item the excessive mass of silicate or AC dust required to fit the late-time spectrum
  \item the need for cold accretion to grow the grains from $\sim 10^{-3}$ to $\sim 0.5$~\msun
  \item the absence of SN-condensed silicate dust
\end{enumerate}

The results presented in this paper are significantly different from the conclusion reached by \cite{ercolano07a} and \cite{wesson15}, who firmly ruled out the possible presence of more than $\sim 3\times 10^{-3}$~\msun\ of dust at the early epochs before $\sim$~day 1200. The main difference between our respective models is that their models were constructed to simultaneously fit the observed UVOIR as well as the IR dust emission. Without this constraint, our models can fit the mid- to far-IR emission at early times with dust having higher temperatures and densities, being optically thick through far-IR wavelengths. At the high densities prevailing at early epochs, the dust is thermally coupled to the gas, and more likely to be collisionally heated by the ambient hot gas, rather than radiatively heated by UVO photons. In either case, the ultimate heating sources are the decaying radioactivities in the ejecta. This scenario seems to be confirmed by the fact that the late time IR spectra, when the ejecta is optically thin, were fitted with a very narrow range in dust temperatures. Such temperature distributions can be more readily explained if the dust is collisionally heated in the ejecta. In our scenario, and UVOIR emission solely arises from the sources in the interclump region, where the optical depth is very low. Only $\sim 10$\% of the energy is emitted at these wavelengths after the dust formation epoch.

The IR observations provide a robust lower limit of $\sim 1.3\times 10^{-4}$~\msun\ on the mass of radioactive $^{44}$Ti that has formed in the explosion, with a more likely value of $1.6\times 10^{-4}$~\msun, depending on the contribution of external sources in heating the dust.

The proposed scenario requires the very rapid evolution of dust in the SN ejecta. Theoretical calculations show that only $\sim3\times 10^{-3}$ of dust can form by day 700 assuming a smooth ejecta, a value that increases to $\sim 0.08$~\msun\ for clumpy ejecta \citep[e.g.][]{sarangi15}. Further increase in the dust mass can be achieved by increasing the density in the clumps, and by the inclusion of molecular and dust-continuum cooling, processes that were ignored during the dust formation epoch. 

Finally, our model cast doubts on  interpretations for the existence of an evolutionary trend in the dust mass from young SNe to older remnants. Infrared observations of young SNe need to carefully determine the origin of the emission, excluding any possible contributions from IR echoes and shocked interstellar dust. An extensive review of the observations of the IR emission from SNe, and their role as dust sources in the early universe was presented by \cite{gall11c}. In particular, Figure~2 of \cite{gall11c} and Figure~4 in \cite{gall14} show that the inferred dust mass in SNe that are less than $\sim 3$~yr old, is between $\sim (0.5-50)\times 10^{-4}$~\msun, whereas the mass of SN condensed dust inferred from observations of young unmixed remnants, such as Cas~A and the Crab Nebula is about 0.1~\msun\ \citep{barlow10,arendt14,gomez12a,temim13}. This ``trend" could be interpreted as an evolutionary one, suggesting the growth of dust grains in the ejecta with time. However, as noted by \cite{gall11c}, the low dust mass in young SNe could also be an observational selection effect, since it is inferred from only near-IR observations, leaving a large part of the dust spectrum unexplored. In this paper we showed that even when the mid-IR region of the spectrum is observed, optical depth effects can result in a large underestimate of the dust mass in the ejecta. 

\noindent
\acknowledgements
We thank Arkaprabha Sarangi for helpful discussions, and the referee for useful criticism that led to the improvement of the manuscript. This work was supported by NASA Astrophysical Data Analysis Program 13-ADAP13-0094.

%==================================
% BIBLIOGRAPHY
%==================================

\clearpage
\bibliographystyle{$HOME/Library/texmf/tex/latex/misc/aastex52/aas.bst}
\bibliography{$HOME/Dropbox/science/00-Bib_Desk/Astro_BIB.bib}

\end{document}